\documentclass[a4paper,english]{article}
\usepackage[T1]{fontenc}
\usepackage[latin1]{inputenc}

\usepackage{RR}
\usepackage{hyperref}

\providecommand{\tabularnewline}{\\}

\usepackage{a4wide}
\usepackage{epsfig}
\usepackage{enumerate}
\usepackage{amsmath}

\usepackage{amsthm}
\theoremstyle{plain}
\newtheorem{prop}{Proposition}
\newtheorem{theorem}{Theorem}
\theoremstyle{definition}
\newtheorem{definition}{Definition}
\theoremstyle{remark}
\newtheorem{remark}{Remark}

\DeclareMathOperator{\conv}{conv}

\usepackage{babel}

\RRdate{31st August 2006}
\RRauthor{
Nicola Yanev \thanks{choby@math.bas.org}  
and Rumen Andonov \thanks[mail-irisa]{prenom.nom@irisa.fr} 
and Philippe Veber \thanksref{mail-irisa}
and Stefan Balev \thanks{stefan.balev@univ-lehavre.fr}
}
\authorhead{Yanev \emph{et al}}
\RRtitle{Approches de relaxation lagrangienne pour la résolution d'une classe de problème d'appariement en bioinformatique}
\RRetitle{Lagrangian Approaches for a class of Matching Problems in Computational Biology}
\titlehead{Lagrangian approaches for matching problems}
\RRnote{}
\RRnote{}
\RRresume{Cet article propose des algorithmes efficaces pour déterminer l'alignement optimal entre une structure et une séquence protéique, problème connu sous le nom de \emph{protein threading}. Nous posons ce problème comme un cas particulier d'appariement. Nous présentons un modèle formel du problème sous la forme d'une famille de graphes, et des programmes en nombre entiers correspondants. Nous étudions dans un premier temps les propriétés de ces modèles, pour ensuite proposer deux approches de relaxation lagrangienne pour la résolution. Enfin, nous montrons, à l'aide de données expérimentales sur des instances réelles, l'efficacité de ces approches.}
\RRabstract{This paper presents efficient algorithms for solving the problem of aligning a protein structure template to a query amino-acid sequence, known as protein threading problem. We consider the problem as a special case of graph matching problem. We give formal graph and integer programming models of the problem. After studying the properties of these models, we propose two kinds of Lagrangian relaxation for solving them. We present experimental results on real life instances showing the efficiency of our approaches.}
\RRmotcle{alignement séquence-structure, complexité, programmation en nombres entiers, relaxation lagrangienne}
\RRkeyword{sequence-structure alignment, complexity, integer programming, Lagrangian relaxation}
\RRprojets{Symbiose}
\RRtheme{\THBio} 
\URRennes  
\begin{document}
\makeRR   

\section{Preliminaries}

\begin{figure}
\begin{center} \psfig{figure=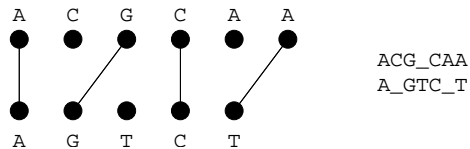, scale=.75}\end{center}

\caption{Matching interpretation of sequence alignment problem}

\label{fig:seqalign}
\end{figure}

Matching is important class of combinatorial optimization problems
with many real-life applications. Matching problems involve choosing
a subset of edges of a graph subject to degree constraints on the
vertices. Many alignment problems arising in computational biology
are special cases of matching in bipartite graphs. In these problems
the vertices of the graph can be nucleotides of a DNA sequence, aminoacids
of a protein sequence or secondary structure elements of a protein
structure. Unlike classical matching problems, alignment problems
have intrinsic order on the graph vertices and this implies extra
constraints on the edges. As an example, Fig.~\ref{fig:seqalign}
shows an alignment of two sequences as a matching in bipartite graph.
We can see that the feasible alignments are 1-matchings without crossing
edges.

\begin{figure}
\begin{center} \psfig{figure=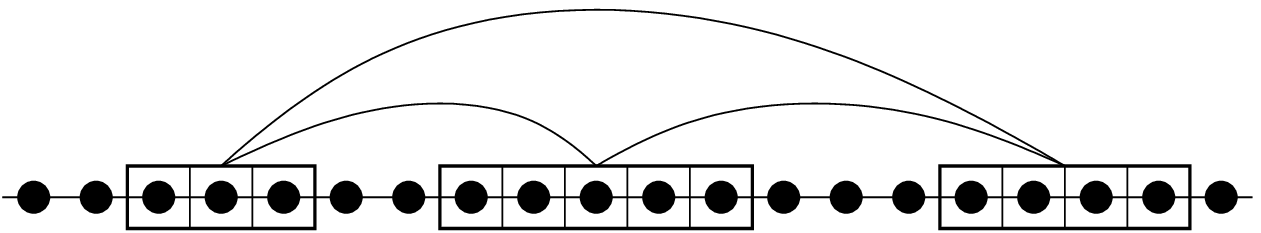, scale=.75}\end{center}

\begin{center}(a)\end{center}

\begin{center}\vskip1.2ex\end{center}


\begin{center}
\footnotesize \setlength{\tabcolsep}{2pt}
\begin{tabular}{|l|rrrrrrrrrrrrrrrrrrrr|}
\hline 
abs.\ position &
 1 &
 2 &
 3 &
 4 &
 5 &
 6 &
 7 &
 8 &
 9 &
 10 &
 11 &
 12 &
 13 &
 14 &
 15 &
 16 &
 17 &
 18 &
 19 &
 20\tabularnewline
\hline
 rel.\ position block 1&
  1&
  2&
  3&
  4&
  5&
  6&
  7&
  8&
  9&
&
&
&
&
&
&
&
&
&
&
\tabularnewline
  rel.\ position block 2&
&
&
&
  1&
  2&
  3&
  4&
  5&
  6&
  7&
  8&
  9&
&
&
&
&
&
&
&
\tabularnewline
  rel.\ position block 3&
&
&
&
&
&
&
&
&
  1&
  2&
  3&
  4&
  5&
  6&
  7&
  8&
  9&
&
&
 \tabularnewline
\hline
\end{tabular}\end{center}{\footnotesize \par}

\begin{center}(b)\end{center}

\begin{center}\vskip1.2ex \psfig{figure=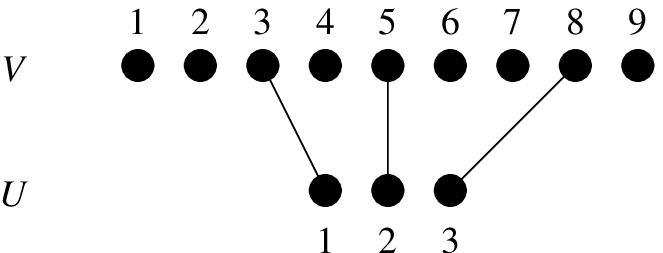, scale=.75}\end{center}

\begin{center}(c) \end{center}

\caption{(a) Example of alignment of query sequence of length 20 and template
containing 3 segments of lengths 3, 5 and 4. (b) Correspondence between
absolute and relative block positions. (c) A matching corresponding
to the alignment of (a).}

\label{fig:threx}
\end{figure}

In this paper we deal with the problem of aligning a protein structure
template to a query protein sequence of length $N$, known as protein
threading problem (PTP). A template is an ordered set of $m$ secondary
structure elements (or blocks) of lengths $l_{i}$, $i=1,\dots,m$.
An alignment (or threading) is covering of contiguous sequence areas
by the blocks. A threading is called feasible if the blocks preserve
their order and do not overlap. A threading is completely determined
by the starting positions of all blocks. For the sake of simplicity
we will use relative positions. If block $i$ starts at the $j$th
query character, its relative position is $r_{i}=j-\sum_{k=1}^{i-1}l_{k}$.
In this way the possible (relative) positions of each segment are
between 1 and $n=N+1-\sum_{i=1}^{m}l_{i}$ (see Fig.~\ref{fig:threx}(b)).
The set of feasible threadings is \begin{equation*}
\mathcal{T} = \{(r_{1},\dots,r_{m}) \;|\; 1 \le r_{1} \le \dots \le r_{m} \le n\}.
\end{equation*}

Protein threading problem is a matching problem in a bipartite graph
$(U\cup V,U\times V)$, where $U=\{ u_{1},\dots,u_{m}\}$ is the ordered
set of blocks and $V=\{ v_{1},\dots,v_{n}\}$ is the ordered set of
relative positions. The threading feasibility conditions can be restated
in terms of matching in the following way. A matching $M\subseteq U\times V$
is feasible if:

\begin{enumerate}[(i)]
\item $d(u)=1$, $u\in U$ (where $d(x)$ is the degree of $x$). This means
that each block is assigned to exactly one position). By the way this
implies that the cardinality of each feasible matching is $m$. 
\item There are no crossing edges, or more precisely, if $(u_{i},v_{j})\in M$,
$(u_{k},v_{l})\in M$ and $i<k$, then $j\leq l$. This means that
the blocks preserve their order and do not overlap. The last inequality
is not strict because of using relative positions. 
\end{enumerate}
Note that while (i) is a classical matching constraint, (ii) is specific
for the alignment problems and makes them more difficult. Fig.~\ref{fig:threx}(c)
shows a matching corresponding to a feasible threading.

\begin{prop}
The number of feasible threadings is $|\mathcal{T}|=\tbinom{m+n-1}{m}$.
\end{prop}

\begin{proof}
We can define the relative positions as $r_{i}=j-\sum_{k=1}^{i-1}l_{k}+i-1$.
In this case the relative positions of the feasible threadings are
related by 
\begin{equation*}
1 \le r_{1} < \dots < r_{m} \le m + n - 1
\end{equation*} 
and a threading is determined by choosing $m$ out of $m+n-1$ positions.
\end{proof}

One of the possible ways to deal with alignment problems is to try
to adapt the existing matching techniques to the new edge constraints
of type (ii). Instead of doing this we propose a new graph model and we develop efficient matching algorithms based on this model.

\begin{figure}
\begin{center} \epsfig{figure=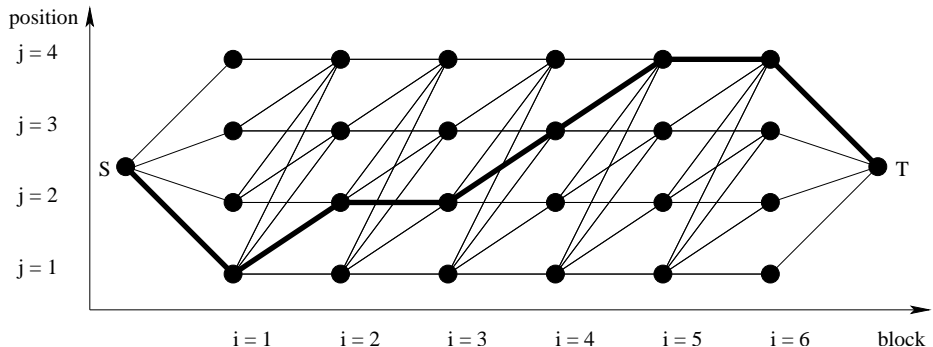, scale=.75}\end{center}

\caption{Example of alignment graph. The path in thick lines corresponds to
the threading in which the positions of the blocks are 1,2,2,3,4,4.}

\label{fig:path}
\end{figure}

We introduce an \emph{alignment graph} $G=(U\times V,E)$. Each vertex
of this graph corresponds to an edge of the matching graph. For simplicity
we will denote the vertices by $v_{ij}$, $i=1,\dots,m$, $j=1,\dots,n$
and draw them as an $n\times m$ grid (see Fig.~\ref{fig:path}).
The vertices $v_{ij}$, $j=1,\dots,n$ will be called $i$th layer. A layer corresponds to a block and each vertex in a layer corresponds to positioning of this block in the query sequence.

One can connect by edges the pairs of vertices of $G$ which correspond
to pairs of noncrossing edges in the matching graph. In this case
a feasible threading is an $m$-clique in $G$. A similar approach
is used in \cite{Lancia_04}. We introduce only a subset of the above edges,
namely the ones that connect vertices from adjacent columns and have
the following regular pattern: $E=\{(v_{ij},v_{i+1,l})\;|\; i=1,\dots,m-1,1\leq j\leq l\leq n\}$.
We add two more vertices $S$ and $T$ and edges connecting $S$ to
all vertices from the first column and $T$ to all vertices from the
last column. Now it is easy to see the one-to-one correspondence between
the set of feasible threadings (or matchings) and the set of $S$-$T$
paths in $G$. Fig.~\ref{fig:path} illustrates this correspondence.

Till now we gave several alternative ways to describe the feasible
alignments. Alignment problems in computational biology involve choosing
the best of them based on some score function. The simplest score
functions associate weights to the edges of the matching graph. For
example, this is the case of sequence alignment problems. By introducing
alignment graphs similar to the above, classical sequence alignment
algorithms, such as Smith-Waterman or Needleman-Wunch, can be viewed
as finding shortest $S$-$T$ paths. When the score functions use
structural information, the problems are more difficult and the shortest
path model cannot incorporate this information.

The score functions in PTP evaluate the degree of compatibility between
the sequence amino acids and their positions in the template blocks.
The interactions (or links) between the template blocks are described
by the so-called generalized contact map graph, whose vertices are
the blocks and whose edges connect pairs of interacting blocks. Let
$L$ be the set of these edges: \begin{equation*}
L = \{(i,k) \;|\; i < k \text{ and blocks \(i\) and \(k\) interact}\}
\end{equation*} Sometimes we need to distinguish the links between adjacent blocks
and the other links. Let $R=\{(i,k)\;|\;(i,k)\in L,\; k-i>1\}$ be
the set of remote (or non-local) links. The links from $L\setminus R$
are called local links. Without loss of generality we can suppose
that all pairs of adjacent blocks interact.

The links between the blocks generate scores which depend on the block
positions. In this way a score function of PTP can be presented by
the following sets of coefficients 

\begin{itemize}
\item $c_{ij}$, $i=1,\dots,m$, $j=1,\dots,n$, the score of putting block
$i$ on position $j$
\item $d_{ijkl}$, $(i,k)\in L$, $1\leq j\leq l\leq n$, the score generated
by the interaction between blocks $i$ and $k$ when block $i$ is
on position $j$ and block $k$ is on position $l$. 
\end{itemize}
The coefficients $c_{ij}$ are some function (usually sum) of the
preferences of each query amino acid placed in block $i$ for occupying
its assigned position, as well as the scores of pairwise interactions
between amino acids belonging to block $i$. The coefficients $d_{ijkl}$
include the scores of interactions between pairs of amino acids belonging
to blocks $i$ and $j$. Loops (sequences between adjacent blocks) may also have sequence specific scores,
included in the coefficients $d_{i,j,i+1,l}$.

The score of a threading is the sum of the corresponding score coefficients
and PTP is the optimization problem of finding the threading of minimum
score. If there are no remote links (if $R=\emptyset$) we can put
the score coefficients on the vertices and the edges of the alignment
graph and PTP is equivalent to the problem of finding the shortest
$S$-$T$ path. In order to take the remote links into account, we
add to the alignment graph the edges 
\begin{equation*}
\{(v_{ij},v_{kl}) \;|\; (i,k) \in R,\; 1 \le j \le l \le n\}
\end{equation*} 
which we will refer as $z$-edges.

\begin{figure}
\begin{center} \epsfig{figure=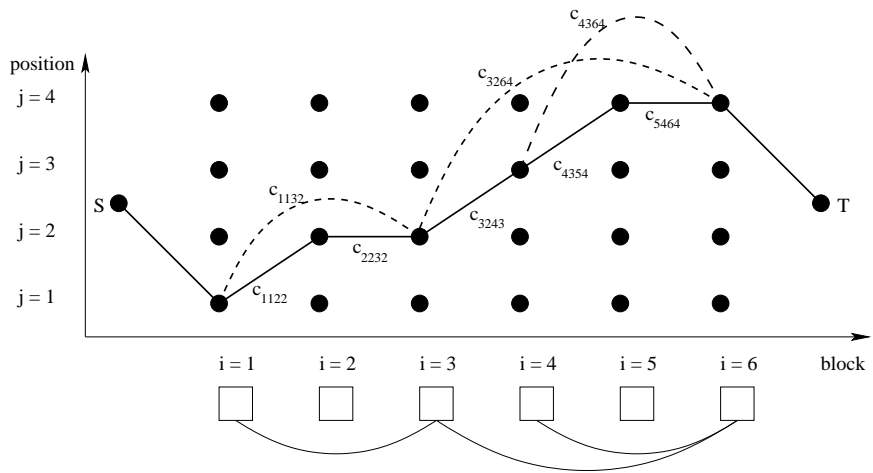, scale=.7}\end{center}

\caption{Example of augmented path. The generalized contact map graph is given
in the bottom. The $x$ arcs of the $S$-$T$ path are in solid lines.
The activated $z$-arcs are in dashed lines. The length of the augmented
path is equal to the score of the threading $(1,2,2,3,4,4)$.}

\label{fig:aug}
\end{figure}

An $S$-$T$ path is said to activate the $z$-edges that have both
ends on this path. Each $S$-$T$ path activates exactly $|R|$ $z$-edges,
one for each link in $R$. The subgraph induced by the edges of an
$S$-$T$ path and the activated $z$-edges is called augmented path.
Thus PTP is equivalent to finding the shortest augmented path in the
alignment graph (see Fig.~\ref{fig:aug}).

As we will see later, the main advantage of this graph is that some
simple alignment problems reduce to finding the shortest $S$-$T$
path in it with some prices associated to the edges and/or vertices.
The last problem can be easily solved by a trivial dynamic programming
algorithm of complexity $O(mn^{2})$. In order to address the general
case we need to represent this graph optimisation problem as an
integer programming problem.

\section{Integer programming formulation}\label{ip}

Let $y_{ij}$ be binary variables associated to the vertices of $G$.
$y_{ij}$ is one if block $i$ is on position $j$ and zero otherwise.
Let $Y$ be the polytope defined by the following constraints:
\begin{align}
  &\sum_{j=1}^{n} y_{ij}  = 1 && i = 1,\dots,m \label{eq:y-assign}\\
  &\sum_{l=1}^{j} y_{il} - \sum_{l=1}^{j} y_{i+1,l} \ge 0 &&
    i = 1,\dots,m-1,\; j = 1,\dots,n-1 \label{eq:y-order}\\
  &y_{ij} \ge 0 && i = 1,\dots,m,\; j = 1,\dots,n \label{eq:y-pos}
\end{align}
Constraints \eqref{eq:y-assign} ensure the feasibility condition (i)
and \eqref{eq:y-order} are responsible for (ii). That is why $Y\cap B^{mn}$
is exactly the set of feasible threadings.

In order to take into account the interaction costs, we introduce a second set of binary variables $z_{ijkl}$, $(i,k) \in L$, $1 \le j \le l \le n$. To avoid added notation we will use vector notation for the variables $y_{i}=(y_{i1},...y_{in})\in B^{n}$ with assigned costs $c_{i}=(c_{i1},...c_{in})\in R^{n}$ and $z_{ik}=(z_{i1k1},\ldots,z_{i1kn},z_{i2k2},\ldots,z_{i2kn},\ldots,z_{inkn})\in B^{\frac{n(n+1)}{2}}$
for $(i,k)\in L$ with assigned costs $d_{ik}=(d_{i1k1},\ldots,d_{i1kn},d_{i2k2},\ldots,d_{i2kn},\ldots,d_{inkn})\in R^{\frac{n(n+1)}{2}}$.

Consider the $2n\times\frac{n(n+1)}{2}$ node-edge incidence matrix of the subgraph spanned by two interacting layers $i$ and $k$. The submatrix $A'$ containing the first $n$ rows (resp. $A''$ containg the last $n$ rows) corresponds to the layer $i$ (resp. layer $k$).

Now the protein threading problem can be defined as
\begin{align}
  &z_{IP}^{L}=v(PTP(L))=\min \{ \sum _{i=1}^{m} c_{i}y_{i} + \sum _{(i,k) \in L}
  d_{ik}z_{ik}\}\label{eq:obj-func}
\end{align}

\vspace*{-0.4cm}
 \begin{align}
\mbox{subject to:}~~  &y=(y_{1},\dots,y_{m})\in Y,  \label{eq:threading-def} \\
  &y_{i}= A' z_{ik} && (i,k)\in L \label{eq:yz-left} \\
  &y_{k}=A'' z_{ik}  &&(i,k)\in L \label{eq:yz-right} \\
  &z_{ik} \in B^{\frac{n(n+1)}{2}} && (i,k)\in L \label{eq:z-plus}
\end{align}
The shortcut notation $v(.)$ will be used for the optimal objective
function value of a subproblem obtained from $PTP(L)$ with some $z$
variables fixed.

\section{Complexity results}\label{complexity}

In this section we study the structure of the polytope defined by (\ref{eq:threading-def})-(\ref{eq:yz-right})
and $z_{ik}\in R_{+}^{\frac{n(n+1)}{2}}$, as well as the impact of the
set $L$ on the complexity of the algorithms for solving the $PTP$
problem. Throughout this section, vertex costs $c_{i}$ are assumed
to be zero. This assumption is not restrictive because the costs $c_{ij}$ can be added to $d_{i,j,i+1,l}$, $l=j,\dots,n$. We will consider the costs $d_{ik}$ as $n \times n$ matrices containing the coefficients $d_{ijkl}$ above the main diagonal and arbitrary large numbers below the main diagonal. In order to simplify the descriptions of the algorithms given in this section we introduce the following matrix operations.

\begin{definition}
Let $A$ and $B$ be two matrices of compatible size. $A \cdot B$ is the matrix product of $A$ and $B$ where the addition operation is replaced by ``$\min$'' and the multiplication operation is replaced by ``+''.
\end{definition}

\begin{definition} \label{def:matmult}
  Let $A$ and $B$ be two matrices of size \(n\times n\).
  \(M = A \otimes B\) is defined by 
  \(M(i,j) = \min_{i \leq r \leq j} A(i,r) + B(i,j)\)
\end{definition}

Below we present four kinds of contact graphs that make PTP polynomially solvable.

\subsection{Contact graph contains only local edges}

As mentioned above, in this case PTP reduces to finding the shortest $S$-$T$ path in the alignment graph which can be done by $O(mn^2)$ dynamic programming algorithm. An important property of an alignment graph containing only local edges is that it has a tight LP description.

\begin{theorem}\label{polytopeY}
The polytope $Y$ is integral, i.e.\ it has only integer-valued vertices.
\end{theorem}

\begin{proof}
Let $A$ be the matrix of the coefficients in \eqref{eq:y-assign}-\eqref{eq:y-order} with columns numbered by the indices of the variables. 
One can prove that $A$ is totaly unimodular (TU) by performing the following sequence of TU preserving transformations.

\medskip
\noindent
\hspace*{1em} \textbf{for} $i=1,\dots,n$\\
\hspace*{2em} delete column $(i,n)$ (these are unit columns)\\
\hspace*{1em} \textbf{for} $i=1,\dots,m$\\
\hspace*{2em} \textbf{for} $j=n-1,\dots,1$\\
\hspace*{3em} pivot on $a_{ij}$ ($A$ is TU iff the matrix obtained by a pivot operation on $A$ is TU\\
\hspace*{3em} delete column $(i,j)$ (now this is unit column)
\medskip

The final matrix is an unit column that is TU. Since all the transformations are TU preserving, $A$ is TU and $Y$ is integral.

One could prove the same assertion by showing that
an arbitrary feasible solution to (\ref{eq:y-assign})-(\ref{eq:y-pos})
is a convex combination of some integer-valued vertices of $Y$. The
best such vertex (in the sense of an objective function) might be
a good approximate solution to a problem whose feasible set is an
intersection of $Y$ with additional constraints.

Let $y$ is an arbitrary non -integer solution to (\ref{eq:y-assign})-(\ref{eq:y-pos}).
Because of (\ref{eq:y-assign}), (\ref{eq:y-order}) an unit flow%
\footnote{The 4 indeces $i,k,p,j\;$used for arcs labeling follows the convention:
tail at vertex $(i,k)$$\;$head at vertex $(p,j)$. Sometimes the
brackets will be dropped.%
} $f=(f_{sj},\; f_{(i,k)(i+1,j)})\;\; i=1,m-1\;\; j=1,n\;\;$$\;$in $G$
exist s.t.\[
\sum_{k\leq j}f_{(i,k)(i+1,j)}=y_{ij}\;\; i=1,m-1\;\; f_{sj}=y_{1j}\;\; j=1,n\]

By the well known properties of the network flow polytope, the flow
$f$ can be expressed as a convex combination of integer-valued unit
flows (paths in $G)$. But each such flow corresponds to an integer-valued
$y$, i.e.$\;$$y_{ij}=f_{(i-1,k)(ij)}=1$. Thus, the convex combination
of the paths that gives $f$ is equivalent to a convex combination
of the respective vertices of $Y$ that gives $y$.

The details for efficiently finding of the set of the vertices participating
in the convex combination could be easily stressed by this sketch
of the prove.
\end{proof}

\subsection{Contact graph contains no crossing edges}

Two links $(i_{1},k_{1})$ and $(i_{2},k_{2})$ such that $i_{1}<i_{2}$
are said to be crossing when $k_{1}$ is in the open interval $(i_{2},k_{2})$.
The case when the contact graph $L$ contains no crossing edges has
been mentioned to be polynomially solvable for the first time in \cite{Aku99}.
Here we present a different sketch for $O(mn^{3})$ complexity of PTP
in this case.

If $L$ contains no crossing edges, then $PTP(L)$ can be recursively
divided into independent subproblems. Each of them consists in computing
all shortest paths between the vertices of two layers $i$ and $k$,
discarding links that are not included in $(i,k)$. The result
of this computation is a distance matrix $D_{ik}$ such that $D_{ik}(j,l)$
is the optimal length between vertices $(i,j)$ and $(k,l)$. Note
that for $j>l$, as there is no path in the graph, $D_{ik}(j,l)$
is an arbitrarily large coefficient. Finally, the solution of $PTP(L)$
is the smallest entry of $D_{1m}$.

We say that a link $(i,k),i<k$ is included in the interval $[a,b]$
when $[i,k]\subseteq[a,b]$. Let us denote by $L_{(ik)}$ the set
of links of $L$ included in $[i,k]$. Then, an algorithm to compute
$D_{ik}$ can be sketched as follows: 

\begin{enumerate}
\item If $L_{(ik)}=\{(i,k)\}$ then the distance matrix is given by 
\begin{equation}
D_{ik}=\left\{ \begin{array}{ll}
d_{ik} & \mbox{if}(i,k)\in L\\
\tilde{0} & \mbox{otherwise}\end{array}\right.
\end{equation}
where $\tilde{0}$ is an upper triangular matrix in the previously
defined sense (arbitrary large coefficients below the main diagonal)
and having only zeros in its upper part.

\item Otherwise, as $L_{(ik)}$ has no crossing edges, there exists some
$s\in[i,k]$ such that any edge of $L_{(ik)}$ except $(i,k)$ is included
either in $[i,s]$ or in $[s,k]$. Then 
\begin{equation}
D_{ik}=\left\{ \begin{array}{ll}
D_{is} \cdot D_{sk}+d_{ik} & \mbox{if}(i,k)\in L\\
D_{is} \cdot D_{sk} & \mbox{otherwise}\end{array}\right.\end{equation} 
\end{enumerate}

If the contact graph has \(m\) vertices, and contains no crossing edges, then the problem is decomposed into \(O(m)\) subproblems.
 For each of them, the computation of the corresponding distance matrix is a \(O(n^{3})\) procedure (matrix multiplication with \((\min,+)\) operations).
 Overall complexity is thus \(O(mn^{3})\). Typically, \(n\) is one or two orders of magnitude greater than \(m\), and in practice,
 this special case is already expensive to solve.

\subsection{Contact graph is a single star}

\label{sec:star}

A set of edges 
$L_{(i)}=\{(i,k_{1}),\dots,(i,k_{r})\}$, $k_{1}<k_{2}<\ldots k_{r}$
is called a \emph{star}%
\footnote{This definition corresponds to the case when all edges have their
left end tied to a common vertex. Star can be symmetrically defined:
i.e. all edges have their right end tied to a common vertex. All proofs
require minor modification to fit this case.%
}.

\begin{theorem}\label{th:starsol} 
Let \(L_{(i)} = \{ (i,k_{1}),\dots,(i,k_{r}) \}\) be a star.
\noindent Then \(D_{ik_{r}} = (\dots(d_{ik_{1}} \otimes d_{ik_{2}}) \otimes \dots) \otimes d_{ik_{r}}\).
\end{theorem} 

\begin{proof}
 The proof follows the basic dynamic programming recursion for this
 particular case: for the star \(L= \{(i,k_{1}),\dots,(i,k_{r})\} = L' \bigcup \{ (i,k_{r}) \}\), 
we have \(v(L:z_{ijk_{r}l}=1) = d_{ijk_{r}l} + \min_{j \leq s \leq l} v(L':z_{ijk_{r-1}s}=1)\).
\end{proof}

In order to compute $A\otimes B$, we use the following recursion:
let $M'$ be the matrix defined by $M'(i,j)=\min_{i\leq r\leq j}A(i,r)$,
then \[
M'(i,j)=\min\{ M'(i,j-1),A(i,j)\},\mbox{ for all }j\geq i\]
 Finally $A\otimes B=M'+B$. From this it is clear that $\otimes$
multiplication for $n\times n$ matrices is of complexity $O(n^{2})$ and hence the complexity of PTP in this case is $O(rn^2)$.

\subsection{Contact graph is decomposable} \label{sec:seq} 

Given a contact graph $L=\{(i_{1},k_{1}),\dots,(i_{r},k_{r})\}$,
$PTP(L)$ can be decomposed into two independent subproblems when
there exists an integer $e\in(1,m)$ such that any edge of $L$ is
included either in $[1,e]$, either in $[e,m]$. Let $I=\{ i_{1},\dots,i_{s}\}$
be an ordered set of indices, such that any element of $I$ allows
for a decomposition of $PTP(L)$ into two independent subproblems.
Suppose additionally that for all $t\leq s-1$, one is able to compute
$D_{i_{t}i_{t+1}}$. Then we have the following theorem:

\begin{theorem}\label{th:sequenceshortestpath}
Let 
$p=(p_{1},p_{2},\dots,p_{n}) = 
D_{i_{1}i_{2}} \cdot D_{i_{2}i_{3}} \cdot \ldots \cdot D_{i_{s-1}i_{s}} \cdot \overline{p}$, where \(\overline{p}=(0,0,\dots,0)\).
 Then for all \(i\), \(p_{i} = v(PTP(L:y_{1i} = 1))\), and \(v(PTP(L))=\min_{1\le i \le n}\{p_{i}\}\).
\end{theorem} 

\begin{proof}
Each multiplication by \(D_{i_{k}i_{k+1}}\) in the definition of \(p\) is an algebraic restatement of 
the main step of the algorithm for solving the shortest path problem in a graph without circuits.
\end{proof} 

With the notations introduced above, the complexity of \(PTP(L)\) for a sequence of such subproblems is \(O(sn^{2})\) plus the cost of computing matrices \(D_{i_{t}i_{t+1}}\).

From the last two special cases, it can be seen that if the
contact graph can be decomposed into independent subsets, and if these
subsets are single edges or stars, then there is a $O(srn^{2})$ algorithm,
where $s$ is the cardinality of the decomposition, and $r$ the maximal
cardinality of each subset, that solves the corresponding PTP.

\begin{remark}\label{rem:no_star_tu}
As a corollary from theorem~\ref{polytopeY} we can easily derive that when
 \(L\) is cross free and does not contain stars, the polytope defined by 
(\ref{eq:yz-left})-(\ref{eq:yz-right}) and \(z_{ik} \in R_{+}^{\frac{n(n+1)}{2}}\)
is  integer.
\end{remark}

\subsection{The threading polytope}

Let $P_{yz}$ be the polytope defined by \eqref{eq:threading-def}-\eqref{eq:yz-right} and $z_{ik} \in R_+^{\frac{n(n+1)}{2}}$ and let $P_{yz}^I$ be the convex hull of the feasible points of \eqref{eq:threading-def}-\eqref{eq:z-plus}. We will call $P_{yz}^I$ a threading polytope.

All of the preceeding polynomiality results were derived without any
refering to the LP relaxation of (\ref{eq:obj-func})-(\ref{eq:z-plus}). The reason is that even for a rather simple version of the graph $L$ the polytope $P_{yz}$ is non-integral. 
We have already seen (indirectly) that if $L$ contains only local links
then $P_{yz} = P^I_{yz}$. Recall the one-to-one
correspondence 
between the threadings, defined as points in $Y$ and the paths in
graph $G$. If $L=\{(i,i+1),i=1,m-1\}$ 
then  $P_{yz}$ is a linear
description of a unit flow in $G$ that is an integral
polytope. Unfortunately, this happenens to be  
a necessary condition also.

\begin{theorem}Let $n\geq 3$ and $L$ contains all local links. Then
$P^I_{yz}=P_{yz}$  if and only if
$R=\textrm{Ø}$.
\end{theorem}

\begin{proof}
$(\Rightarrow)$
Without loss of generality we can take $R=(1,3)$, $m=3$ and $n=3$. Then
the point $\; A=(y_{11}=y_{12}=y_{21}=y_{22}=0.5,y_{32}=0.75,\; y_{33}=0.25,\; z_{1121}=z_{2132}=z_{1222}=z_{1232}=0.5,z_{2232}=z_{2233}=z_{1132}=z_{1133}=0.25 )\;\in P_{yz}\;$
and the only eligible (whose convex hull could possibly contain $A$)$\;$integer-valued
vertices of $P_{yz}\;$ are $B=(y_{11}=y_{21}=y_{32}=z_{1132}=1)$
and $\;C=(y_{12}=y_{22}=y_{32}=z_{1232}=1)\;$ but $A\;$
is not in the segment $[B,C]$. The generalization of this proof for
arbitrary$\;$ $m,n\geq3\;$ and $R$ is almost straighforward.

$(\Leftarrow)$ Follows directly from Theorem~\ref{polytopeY}.
\end{proof}

This is a kind of negative result seting a limit to relying on LP solution.

\section{Lagrangian approaches}\label{lr} 
Consider an integer program 
\begin{equation}
z_{IP}=\min\{ cx:x\in S\},\mbox{where } S=\{ x\in Z_{+}^{n}:Ax\leq b\}\label{ipfor}
\end{equation}
 Relaxation and duality are the two main ways of determining $z_{IP}$
and upper bounds for $z_{IP}$. The linear programming relaxation
is obtained by changing the constraint $x\in Z_{+}^{n}$ in the definition
of $S$ by $x\geq0$. The Lagrangian relaxation is very convenient
for problems where the constraints can be partitioned into a set of
``simple'' ones and a set of ``complicated'' ones. Let us assume for example
that the complicated constraints are given by $A^{1}x\leq b^{1}$,
where $A^{1}$ is $m\times n$ matrix, while the nice constraints
are given by $A^{2}x\leq b^{2}$. Then for any $\lambda\in R_{+}^{m}$
the problem 
$$z_{LR}(\lambda)=\min_{x\in Q}\{ cx+\lambda(b^{1}-A^{1}x)\}$$
where $Q=\{ x\in Z_{+}^{n}:A^{2}x\leq b^{2}\}$ is Lagrangian
relaxation of (\ref{ipfor}), i.e. $z_{LR}(\lambda)\leq z_{IP}$ for
each $\lambda\geq 0$. The best bound can be obtained by solving the
Lagrangian dual $z_{LD}=\max_{\lambda\geq0}z_{LR}(\lambda)$.
It is well known that relations $z_{IP}\geq z_{LD}\geq z_{LP}$ hold.

An even better relaxation, called \emph{cost-splitting}, can be obtained
by applying Lagrangian duality to the reformulation of \eqref{ipfor}
given by 
\begin{align}
  &z_{IP}=\min cx^{1} \label{eq:obj-rjp}
\end{align} 

\begin{align}
\mbox{subject to:}~~  &A^{1}x^{1}\leq b^{1}, ~~ A^{2}x^{2}\leq b^{2},\label{eq:cntsrA} \\
  &x^{1} - x^{2} =0 \label{eq:egality} \\
  &x^{1}\in Z_{+}^{n}, ~~  x^{2}\in Z_{+}^{n},\label{eq:x}
\end{align} Taking $x^{1}-x^{2}=0$ as the complicated constraint, we obtain
the Lagrangian dual of (\ref{eq:obj-rjp})-(\ref{eq:x}) \begin{align}
  &z_{CS}=\max_{u}\{\min c^{1}x^{1} + \min c^{2}x^{2} \}\label{eq:dual-rjp}
\end{align} 

\begin{align}
\mbox{subject to:}~~  &A^{1}x^{1}\leq b^{1}, ~~ A^{2}x^{2}\leq b^{2},\label{eq:dual-cntsrA} \\
  &x^{1}\in Z_{+}^{n}, ~~  x^{2}\in Z_{+}^{n},\label{eq:dual-x}
\end{align} where $u=c^{2},c^{1}=c-u$.

The following well known polyhedral characterization of the cost splitting dual will be used later:
\begin{theorem}[see \cite{NW88}]\label{th:cs_char}
$$
z_{CS} = \max \left\{ cx:\ \conv\{x \in Z_+^n :\ A^1 x \leq b^1\} \cap \conv\{x \in Z_+^n :\ A^2 x \leq b^2\} \right\}
$$
where $\conv\{A\}$ denotes the convex hull of $A$.
\end{theorem}

In both relaxations in order to find $z_{LD}$ or $z_{CS}$ one has
to look for the maximum of a concave piecewise linear function. This
appeals for using the so called subgradient optimization technique. For
the function $z_{LR}(\lambda)$, the vector $s^{t}=b^{1}-A^{1}x^{t}$,
where $x^{t}$ is an optimal solution to ${\min_{Q}\{ cx+\lambda^{t}(b^{1}-A^{1}x)\}}$,
is a subgradient at $\lambda^{t}$. 
The following subgradient algorithm is an analog of the steepest ascent method of maximizing a function:
\begin{itemize}
\item (Initialization): Choose a starting point $\lambda^{0}$, $\Theta_{0}$
and $\rho$. Set $t=0$ and find a subgradient $s^{t}$. 
\item While $s^{t}\neq0$ and $t<t_{\max}$ do \{ $\lambda^{t+1}=\lambda^{t}+\Theta_{t}s^{t}$;
$t\leftarrow t+1$; find $s^{t}$\}
\end{itemize}
This algorithm stops either when $s^{t}=0$, (in which case $\lambda^{t}$
is an optimal solution) or after a fixed number of iterations. We experimented two
schemes for selecting $\{\Theta_{t}\}$:
\begin{align}
\Theta_{t} & = \Theta_{0}\rho^{t}\\
\Theta_t & = \Theta_{0} \frac{ \kappa_t (U_t - L_t) \rho^t}{||s^t||_1}
\end{align}
where 
\begin{align*}
&0<\rho<1\\
&\{ \kappa_t \} \mbox{ is a random sequence whose terms are uniformly chosen in [1, 1.4]}\\
&L_t \mbox{ is the best value of } z_{LR}(\lambda) \mbox{ up to iteration } t\\
&U_t \mbox{ is the best value of any feasible solution found up to iteration } t\\
&||s^t||_1 \mbox{ is the 1-norm of the subgradient}
\end{align*}

\section{Lagrangian relaxation}\label{section:sblr}

Relying on complexity results from section \ref{complexity}, we show
now how to apply Lagrangian relaxation taking as complicating
constraints \eqref{eq:yz-right}. Recall that these constraints insure that the $y$-variables and the $z$-variables select the same position of block $k$. Associating Lagrangian multipliers $\lambda_{ik}$ to the relaxed constraints we obtain

$$
z_{LR}(\lambda) = \min_{y,z} \left\{ \sum _{i=1}^{m} c_{i}(\lambda)y_{i} + \sum _{(i,k) \in L}
  d_{ik}(\lambda)z_{ik} \right\}
$$
where
$$c_i(\lambda) = c_i + \sum_{(k,i) \in L} \lambda_{ki}, \;\;
d_{ik}(\lambda) = \sum_{(i,k) \in L} (d_{ik} - \lambda_{ik} A'')$$

Consider this relaxation for a fixed $\lambda$. Suppose that a block $i$ is on position $j$ in the optimal solution. Then the optimal values of the variables $z_{ijkl}$ can be found using the method desctibed in section~\ref{sec:star}. In this way the relaxed problem decomposes to a set of independent subproblems. Each subproblem has a star as a contact graph. After solving all the subproblems, we can update the costs $c_i(\lambda)$ with the contribution of the star with root $i$ and find the shortest $S$-$T$ path in the alignment graph.

Note that for each $\lambda$ the solution defined by the $y$-variables is feasible to the original problem. In this way at each iteration of the subgradient optimisation we have an heuristic solution. At the end of the optimization we have both lower and upper bounds on the optimal objective value.

Symmetrically, we can relax the left end of each link or even relax the left end of one part of the links and the right end of the rest. The last is the approach used in \cite{balev-2004}. The same paper describes a branch-and-bound algorithm using this Lagrangian relaxation instead of the LP relaxation.

\section{Cost splitting}\label{section:costsplit}

In order to apply the results from the previous sections, we need to
find a suitable partition of $L$ into $L^1\bigcup L^2...\bigcup L^t$
where each $L^s$ induces an easy solvable $PTP(L^s)$, and to use the
cost-splitting variant of the Lagrangian duality. Now we can 
restate (\ref{eq:obj-func})-(\ref{eq:z-plus}) equivalently as:
\begin{align}
  & z_{IP}^L=\min \left\{\sum_{s=1}^t (\sum _{i=1}^m c_i^s y_i^s +
    \sum _{(i,k) \in L^s} d_{ik}z_{ik})\right\}\label{eq:zipL}
\end{align}
\vspace*{-0.4cm}
\begin{align}
\mbox{subject to:}~~  &y_i^1=y_i^s, &&s=2,t\label{eq:complicating}\\
  & y^s=(y_1^s,..y_m^s)\in Y,&& s=1,\dots,t\label{eq:sub-threading-def}\\
  &y_i^s= A_iz_{ik},\:\:y_k^s=A_kz_{ik} && s=1,\dots,t && (i,k)\in L^s \label{eq:sub-yz-left-right}\\
  &z_{ik} \in B^{\frac{n(n+1)}{2}} && s=1,\dots,t && (i,k) \in L^s \label{eq:sub-threading-defz}
\end{align}

Taking (\ref{eq:complicating}) as the complicating constraints, we
obtain the Lagrangian dual of $PTP(L)$:
 \begin{equation}
 z_{CS}=\max_\lambda\min_y\sum_{s=1}^t
    (\sum _{i=1}^m  c_i^s(\lambda)y_i^s +
    \sum _{(i,k) \in L^s} d_{ik}z_{ik})= \max_\lambda\sum_{s=1}^t z_{IP}^{L^s}(\lambda)\label{eq:split-obj}
\end{equation}
subject to (\ref{eq:sub-threading-def}), (\ref{eq:sub-yz-left-right}) and (\ref{eq:sub-threading-defz}).

The Lagrangian multipliers $\lambda^s$ are associated with the
equations (\ref{eq:complicating}) 
and $c_i^1(\lambda)=c_i^1+\sum_{s=2}^t\lambda^s$, $c_i^s(\lambda)=c_i^s-\lambda^s, s=2,\dots,t$. The coefficients $c_i^s$ are arbitrary (but fixed) decomposition (cost-split)
of the coefficients $c_i$, i.e.\ given by $c_i^s=p_sc_i$ with $\sum
p_s=1$. 

From the Lagrangian duality theory it follows that $z_{LP}\le
z_{CS}\le z_{IP}$. 
However choosing the decomposition remains a delicate issue. A
tradeoff has to be found between tightness of the bound and
complexity of the dual. At one extreme,
when decomposing the interaction graph into cross-free sets, the dual
problem is of $O(mn^3)$ complexity. This makes this approach hopeless
for practical situations. At the other extreme, each set in the
decomposition could contain a single edge. This is a very favorable
situation for complexity matters, but it turns out that in this case,
the cost-splitting dual boils down to LP bound:

\begin{theorem}
If $t = |L|$ then $z_{CS} = z_{LP}$
\end{theorem}

\begin{proof}
From Th. \ref{th:cs_char}, we have
$$ 
z_{CS} = \max \left\{ cy + dz :\ \bigcap_{(i,k) \in L} \conv\{y,z \in
Z_+^n :y_i = A_i^k z_{ik} \wedge y_k = A_k^i z_{ik}\} \right\}
$$
However, as underlined in Rem. \ref{rem:no_star_tu}, the set
$$ \{y,z \in
R_+^n :y_i = A_i^k z_{ik} \wedge y_k = A_k^i z_{ik})\} $$
only has integer extremal points, which amounts to say that
$$
\{y,z \in
R_+^n :y_i = A_i^k z_{ik} \} = \conv\{y,z \in
Z_+^n :y_i = A_i^k z_{ik} \wedge y_k = A_k^i z_{ik}\}
$$

The result follows:
$$
z_{CS} = \max \left\{ cy + dz :\ \bigcap_{(i,k) \in L} \{y,z \in
R_+^n :y_i = A_i^k z_{ik} \wedge y_k = A_k^i z_{ik} \} \right\} = z_{LP}
$$

\end{proof} 
 
By applying the
subgradient optimization technique (\cite{NW88}) in order to obtain $z_{CS}$,
one need to solve $t$ problems $v_{IP}^{L^s}(\lambda)$ for each $\lambda$ generated during
the subgradient iterations. As usual, the most time consuming step is $PTP(L^s)$ solving, but we have demonstrated its $O(n^2)$ complexity in the case when $L^s$ is a union of independent stars.

\section{Experimental results}

In this section we present three kinds of experiments. 
First, in subsection \ref{exact_sol}, we show that the 
 branch-and-bound algorithm based on 
the Lagrangian relaxation from section \ref{section:sblr} (BB\_LR)  can be successfully used for solving exactly 
huge PTP instances. In subsection \ref{approx_sol}, we study  the impact 
of the approximated solutions given by different PTP solvers on the quality of 
the prediction. Lastly, in subsection \ref{LR_vs_CS} we  experimentally compare 
the two  relaxations proposed in this paper and show that they have similar performances.


In order to evaluate the performance of our algorithm and to test it on real problems, we integrated it in the structure prediction tool FROST \cite{marin-2002,Frost02a}. 
FROST (Fold Recognition-Oriented Search Tool) is intended to assess
the reliability of fold assignments to a given protein sequence. In our experiments we used its the structure database, containing about 1200 structure templates, as well as its score function. FROST uses a specific procedure to normalize the alignment score and  to evaluate its significance. As the scores are highly dependent on sequence lengths, for each
template of the database this procedure selects 5 groups  of non homologous sequences corresponding to -30\%, -15\%, 0\%, +15\% and +30\% of the
template length. Each group contains about 200 sequences  of equal length.   Each of the about 1000 sequences is aligned to the template. This procedure involves about 1,200,000 alignments and is extremely computationally expensive \cite{dfrost}.  
The values of the score distribution function $F$ in the points 0.25 and 0.75 are approximated by this empirical data. When a ``real'' query is threaded to this template, the raw alignment score $S$ is replaced by the \emph{normalized distance} $NS =\frac{F(.75)-S}{F(.75) - F(.25)}$. Only the value $NS$  is used to evaluate
the relevance of the computed  raw score to the considered distribution.

\subsection{Solving PTP exactly}\label{exact_sol}

To test the efficiency of our algorithm we used the data from 9,136 threadings made in order to compute the distributions of 10 templates. Figure~\ref{fig:rt} presents the running times for these alignments.
The optimal threading was found in less than one minute for all but 34 instances. For 32 of them the optimum was found in less than 4 minutes and only for two instances the optimum was not found in one hour.
However, for these two instances the algorithm produced in one minute a suboptimal solution with a proved objective gap less than 0.1\%. 

\begin{figure}
\begin{center}
\psfig{figure=rt.eps,scale=0.75}
\caption{%
Running times of 9,136 threading instances as a function of the search space size. The experiment is made on 1.8GHz Pentium PC with 512MB RAM
} \label{fig:rt}
\end{center}
\end{figure}

It is interesting to note that for 79\% of the instances the optimal solution was found in the root of the branch-and-bound tree. This means that the Lagrangian relaxation produces a solution which is feasible for the original problem. The same phenomenon was observed in \cite{Xu03,informs-2004} where integer programming models are solved by linear relaxation. However,  the  dedicated algorithm based of the Lagrangian relaxation from section \ref{section:sblr} is  much faster  than a general purpose solver using the linear relaxation. For comparison, solving instances of size of order $10^{38}$ by CPLEX of ILOG solver reported in \cite{informs-2004} takes  more than one hour on a faster than our computer, while instances of that size were solved by LR algorithm  in about 15 seconds.

The use of  BB\_LR made possible to compute the exact score distributions of all templates from the FROST database for the first time \cite{dfrost}. An experiment on about 200 query proteins of known structure shows that using the new algorithm improves not only the running time of the method, but also its quality. When using the exact distributions, the sensitivity of FROST (measured as the percentage of correctly classified queries) is increased by 7\%. Moreover, the quality of the alignments produced by our algorithm (measured as the difference with the VAST alignments) is also about 5\% better compared to the quality of the alignments produced by the heuristic algorithm.

\subsection{Impact of the approximated solution on the quality of the  prediction }\label{approx_sol}

We compared  BB\_LR to two other algorithms used by FROST -- a steepest-descent heuristic (H) and an implementation of the branch-and-bound algorithm from \cite{Lath96} (B). The comparison was made over 952 instances (the sequences threaded to the template 1ASYA when computing its score distribution). Each of the three algorithms was executed with a timeout of 1 minute per instance. We compare the best solutions produced during this period. The results of this comparison are summarized in Table~\ref{tab:comp}. For the smallest instances (the first line of the table) the performance of the three algorithms is similar, but for instances of greater size our algorithm clearly outperforms the other two. It was timed out only for two instances, while B was timed out for all instances. L finds the optimal solution for all but 2 instances, while B finds it for no instance. The algorithm B cannot find the optimal solution for any instance from the fourth and fifth lines of the table even when the timeout is set to 2 hours. The percentage of the optima found by H degenerates when the size of the problem increases. Note however that H is a heuristic algorithm which produces solutions without proof of optimality. Table~\ref{tab:distr} shows the distributions computed by the three algorithms. The distributions produced by H and especially by B are shifted to the right with respect to the real distribution computed by L. This means that for example a query of length 638AA and score 110 will be considered as significantly similar to the template according to the results provided by B, while in fact this score is in the middle of the score distribution.

\begin{table}
\caption{Comparison between three algorithms: branch-and-bound using Lagrangian relaxation (L), heuristic steepest-descent algorithm (H), and branch-and-bound of Lathrop and Smith (B). The results in each row are average of about 200 instances.} \begin{center}
\begin{tabular}{|rrrr|rrr|rrr|}
\hline
query  & $m$ & $n$ & $|T|$ & \multicolumn{3}{|c|}{average time(s)} & \multicolumn{3}{|c|}{opt(\%)} \\
length &     &     &       & L & H & B & L & H & B \\
\hline
342 &  26 &   4 & 3.65e03 & 0.0 &  0.1 &  0.0  &  100 & 99 & 100 \\
416 &  26 &  78 & 1.69e24 & 0.6 & 43.6 & 60.0  &  100 & 63 &   0 \\
490 &  26 & 152 & 1.01e31 & 2.6 & 53.8 & 60.0  &  100 & 45 &   0 \\
564 &  26 & 226 & 1.60e35 & 6.4 & 56.6 & 60.0  &  100 & 40 &   0 \\
638 &  26 & 300 & 1.81e38 & 12.7 & 59.0 & 60.0  &   99 & 31 &   0 \\
\hline
\end{tabular}
\label{tab:comp}
\end{center}
\end{table}

\begin{table}
\caption{Distributions produced by the three algorithms.} \label{tab:distr}
\begin{center}
\begin{tabular}{|r|rrr|rrr|rrr|}
\hline
query  & \multicolumn{3}{|c|}{distribution (L)} & \multicolumn{3}{|c|}{distribution (H)} & 
   \multicolumn{3}{|c|}{distribution (B)}\\
length & $F(.25)$ & $F(.50)$ & $F(.75)$ & $F(.25)$ & $F(.50)$ & $F(.75)$ & $F(.25)$ & $F(.50)$ & $F(.75)$\\
\hline
342 & 790.5 & 832.5 & 877.6 & 790.5 & 832.6 & 877.6 & 790.5 & 832.5 & 877.6\\
416 & 296.4 & 343.3 & 389.5 & 299.2 & 345.4 & 391.7 & 355.2 & 405.5 & 457.7\\
490 & 180.6 & 215.2 & 260.4 & 184.5 & 219.7 & 263.4 & 237.5 & 290.4 & 333.0\\
564 & 122.6 & 150.5 & 181.5 & 126.3 & 157.5 & 187.9 & 183.3 & 239.3 & 283.4\\
638 &  77.1 & 109.1 & 142.7 &  87.6 & 118.5 & 150.0 & 154.5 & 197.0 & 244.6\\
\hline
\end{tabular}
\end{center}
\end{table}

\begin{figure}
\begin{minipage}[t]{.45\textwidth}
~

\epsfig{figure=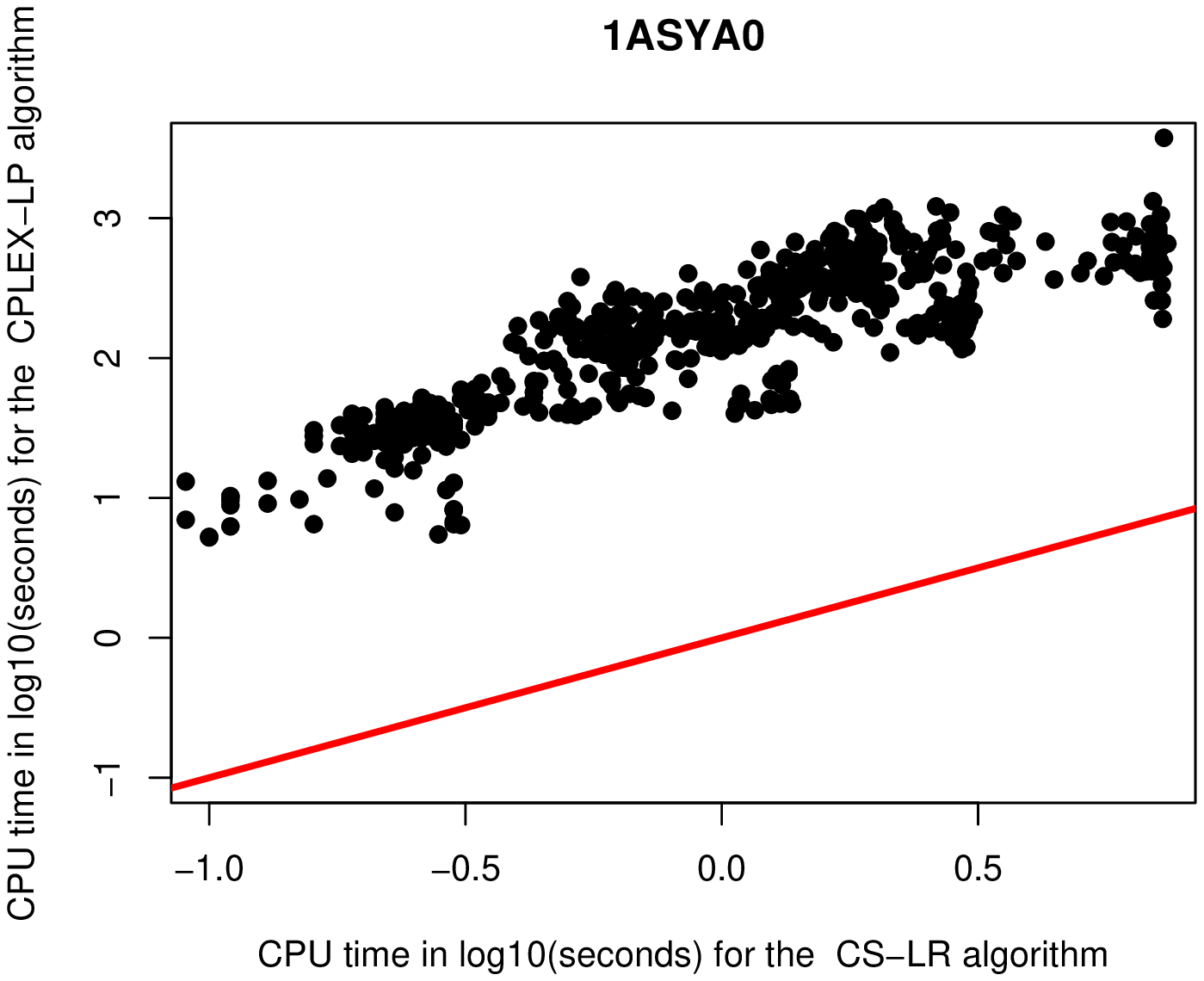, scale=.40}
\end{minipage}
\hfill
\begin{minipage}[t]{.45\textwidth}
 Plot of time in seconds with  CS 
 algorithm
 on the $x$-axis and the LP algorithm from \cite{informs-2004}
on the $y$-axis. Both  algorithms compute  approximated solutions 
for 962 threading instances associated to  the template 
1ASYA0 from the FROST database.  The linear curve in the plot is the 
line $y=x$. What is observed
is a significant  performance gap between the algorithms. For example in 
a point $(x,y)=(0.5,3)$ CS is $10^{2.5}$ times faster than  LP relaxation.
These  results were obtained on 
an Intel(R) Xeon(TM) CPU 2.4~GHz, 2 GB RAM, RedHat 9 Linux.
The MIP models were solved using CPLEX 7.1 solver \cite{ilog}.

\end{minipage}
\caption{Cost-Splitting Relaxation  versus LP Relaxation
\label{times1}}
\end{figure}

We conducted the following experiment. For the purpose 
of this section we chose a set of 12 non-trivial templates. 60 distributions are associated 
to them. We first computed these distributions  
using an exact algorithm for solving the underlying PTP problem. 
The same distributions have been afterwords computed using the approximated 
solutions obtained by any of the three algorithms here considered. 
By  approximated solution we mean respectively the following: i) for a MIP model
this is the solution given by the LP relaxation; ii) for 
the Lagrangian Relaxation  (LR) algorithm this is the solution 
obtained for 500 iterations (the upper bound used  in \cite{balev-2004}). 
Any exit with less than 500 iterations is a sign  that the exact value has been found;
iii) for the Cost-Splitting algorithm   (CS) this is the solution 
obtained either for 300 iterations or when the relative error between upper and lower bound is less than $0.001$. 

We use the MYZ integer programming  model introduced in \cite{informs-2004}. 
It has been proved faster than the MIP model used in 
the package  RAPTOR \cite{Xu03} which was well ranked among all non-meta servers  in 
CAFASP3 (Third Critical Assessment of Fully Automated Structure Prediction) and in
CASP6 (Sixth Critical Assessment of Structure Prediction). 
Because of time limit we  present here the results from 
10 distributions  only\footnote{More data will be solved and provided for the final version.}.  
Concerning the 1st  quartile the relative 
error between the exact and approximated solution 
is $3\times10^{-3}$ in two cases over all 2000 instances 
and  less than $10^{-6}$ for all other cases.  Concerning the 3rd  quartile, the relative 
error is  $10^{-3}$ in two cases and  less than $10^{-6}$ for all other cases.

\begin{figure}	
\vspace*{-1cm}
\begin{minipage}[t]{.45\textwidth}
\epsfig{figure=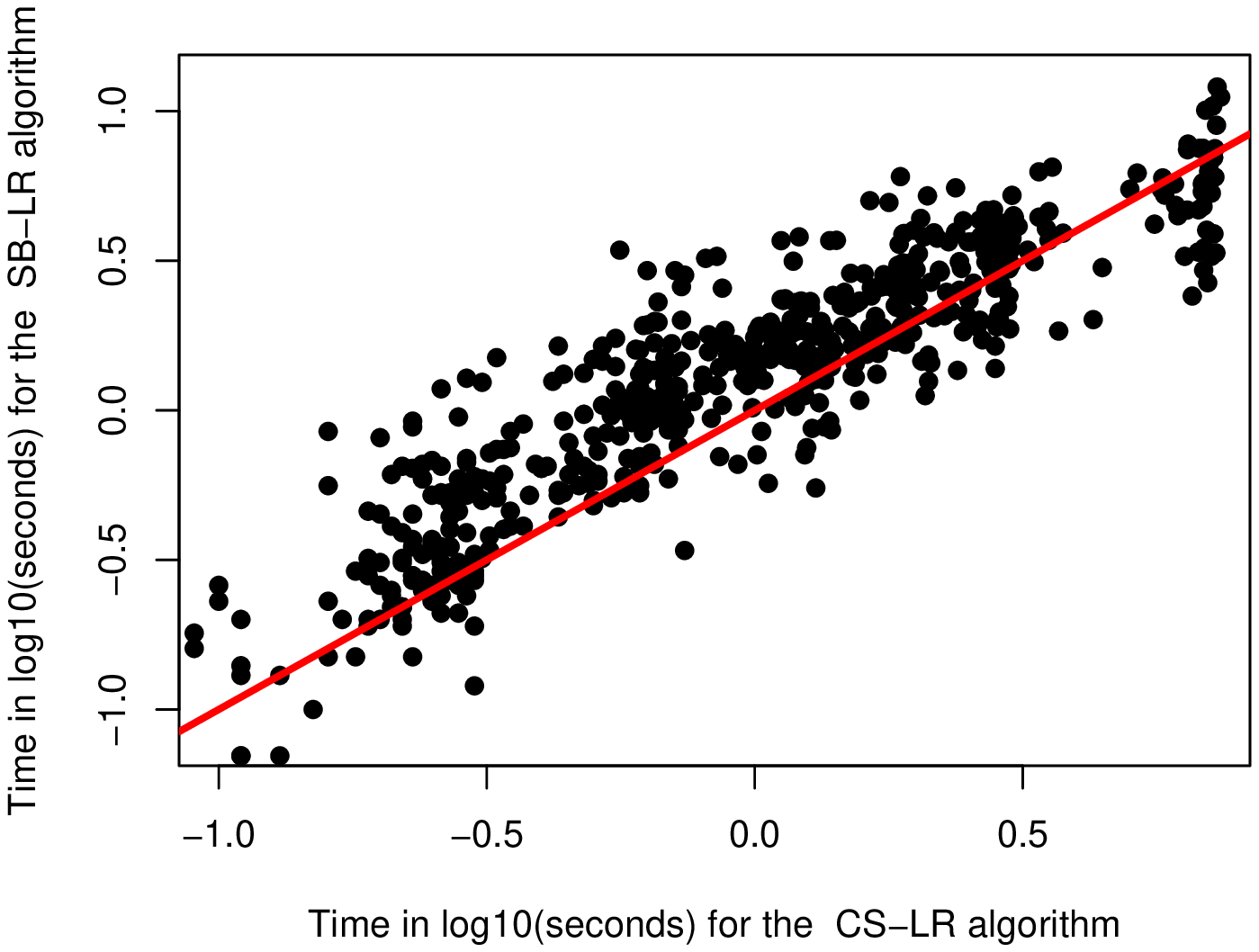, scale=.38}
\end{minipage}
\hfill
\begin{minipage}[t]{.45\textwidth}
\epsfig{figure=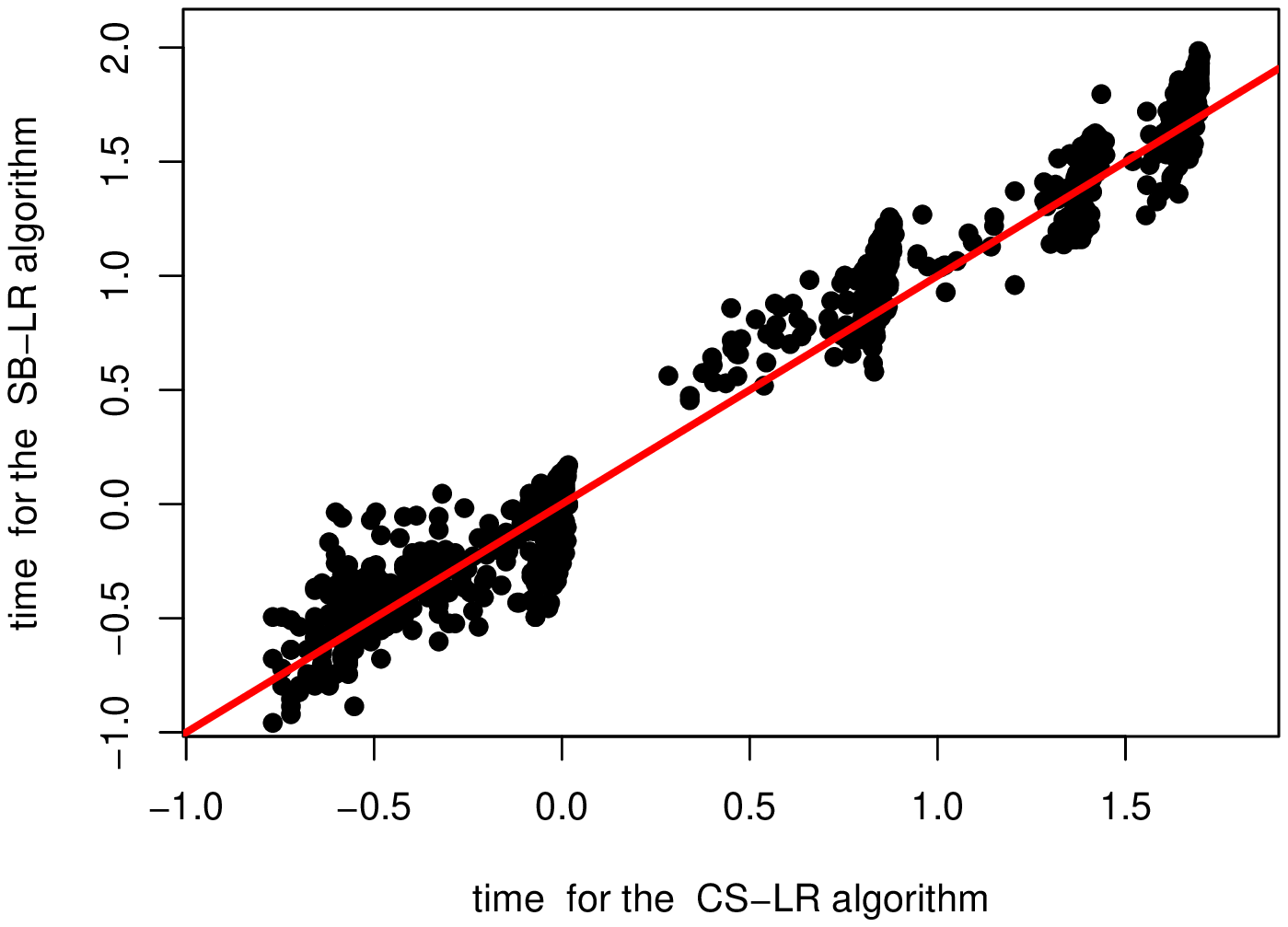, scale=.38}
\end{minipage}
\caption{Plot of time in seconds with  CS
(Cost-Splitting Relaxation) algorithm
 on the $x$-axis versus LR (Lagrangian Relaxation) 
algorithm \cite{balev-2004} on the $y$-axis concerning score distributions 
of two templates. Both the  $x$-axis and 
$y$-axis are in logarithmic scales. The linear curve in the plot is the 
line $y=x$.  \textbf{Left:} The template 1ASYA (the one referenced in
\cite{balev-2004}) has been threaded with 962 sequences. 
\textbf{Right: } 1ALO\_0 is one of the templates yielding the biggest problem instances 
when aligned with the 704  sequences associated to it in the database. 
We observe that although CS is often faster than LR, in general the performance of both algorithms 
is very close. 
\label{times2}}
\vspace*{-0.7cm}
\end{figure}

All 12125 alignments for the set of 60 templates have been computed 
by the other two algorithms. Concerning the 1st  quartile, the 
exact and approximated solution are equal for all cases for both  (LR and CS) algorithms. 
Concerning the 3rd  quartile and in case of  LR  algorithm the 
exact  solution equals the approximated one in all but two cases in which the relative error is
respectively  $10^{-3}$ and $10^{-5}$. In the same quartile and 
in case of  CS  algorithm  the exact  solution equals the approximated one in 
12119 instances and the relative error is  $7\times10^{-4}$ in only 6 cases.  

Obviously, this loss of precision (due to computing the distribution by not always taking 
the optimal solution) is negligible  and does not  degrade the 
quality of  the prediction. We therefore conclude that the approximated solutions 
given by any of above mentioned algorithms can be successfully used in the 
score distributions phase. 

\subsection{ Cost splitting versus Linear Programming and Lagrangian relaxations}\label{LR_vs_CS}

\begin{figure}[t]
\begin{minipage}[t]{.45\textwidth}
~

\epsfig{figure=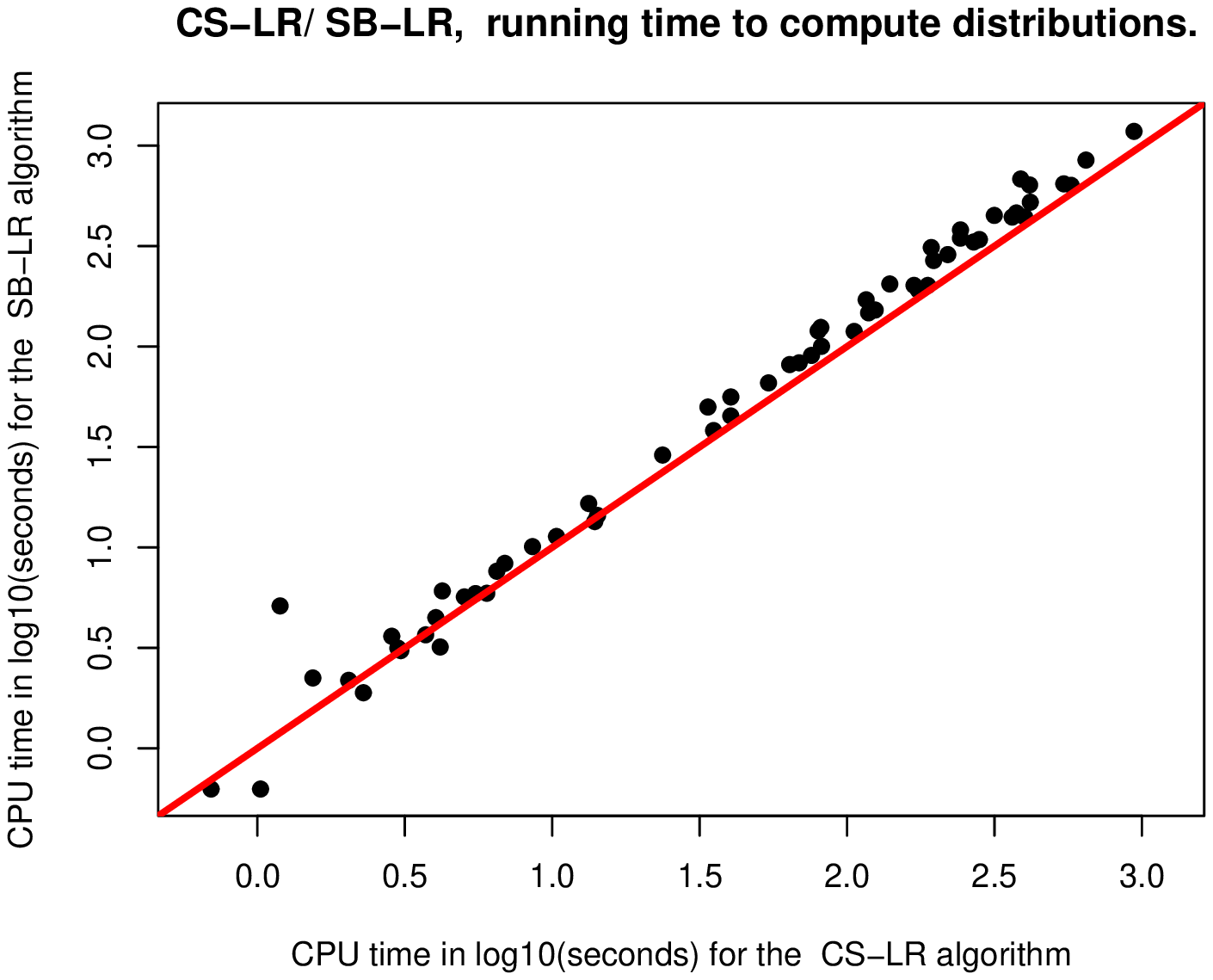, scale=.40}
\end{minipage}
\hfill
\begin{minipage}[t]{.45\textwidth}
~ 

 Plot of time in seconds with  CS  algorithm
 on the $x$-axis and the LR algorithm 
on the $y$-axis. Each point corresponds to the total time needed to compute one 
distribution determined by approximately 200 alignments of the same size.  
61 distributions have been computed 
which needed solving totally 12125 alignments.   Both the  $x$-axis and 
$y$-axis are in logarithmic scales. The linear curve in the plot is the line $y=x$.
CS  is consistently faster than the LR algorithm.
\end{minipage}
\caption{CS versus LR : recapitulation plot concerning 12125 alignments. \label{times3}}
\end{figure}

Our third  numerical experiment concerns running time comparisons for computing 
approximated solutions by LP, LR and CS algorithms. The obtained results are summarized on 
figures \ref{times1}, \ref{times2} and \ref{times3}. Figure \ref{times1} clearly shows that 
CS algorithm significantly outperforms the LP relaxation.  Figures  \ref{times2} and \ref{times3} 
compare CS with LR algorithm and illustrate that they give close running times 
(CS being slightly faster than LR).  
Time sensitivity with respect to the size of the problem is given in Fig. \ref{times3}.

\section{Conclusion}

The results presented in this paper confirm once more that integer programming approach is well suited to solve the protein threading problem. Even if the possibilities of general purpose solvers using linear programming relaxation are limited to instances of relatively small size, one can use the specific properties of the problem and develop efficient special purpose solvers. After studying these properties we propose two Lagrangian approaches, Lagrangian relaxation and cost splitting. These approaches are more powerful than the general integer programming and allow to solve huge instances\footnote{Solution space size of  $10^{40}$ corresponds to a MIP model with $4 \times 10^{4}$
 constraints and $2\times 10^{6}$ variables \cite{yanev-2004}.}, with solution space of size up to $10^{77}$, within a few minutes. 

The results lead us to think that even better performance could be obtained by relaxing additional constraints, relying on the quality of LP bounds. In this manner, the relaxed problem  will be easier to solve. This is the subject of our current work.

This paper deals with the problem of global alignment of protein sequence and structure template. But the methods presented here can be adapted to other classes of matching problems arising in computational biology. Examples of such classes are semi-global alignment, where the structure is aligned to a part of the sequence (the case of multi-domain proteins), or local alignment, where a part of the structure is aligned to a part of the sequence. Problems of structure-structure comparison, for example contact map overlap, are also matching problems that can be treated with similar techniques. Solving these problems by Lagrangian approaches is work in progress.

\end{document}